\documentclass[a4paper]{jpconf}

\usepackage{graphicx}
\usepackage{dcolumn}
\usepackage{bm}
\usepackage{ifpdf}

\begin{document}
\title{Charged black hole solutions of non-linear electrodynamics and generalized gauge field theories}

\author{J. Diaz-Alonso, D. Rubiera-Garcia}

\address{LUTH, Observatoire de Paris, CNRS, Universit\'e Paris
Diderot. 5 Place Jules Janssen, 92190 Meudon, France, and}
\address{Departamento de Fisica, Universidad de Oviedo. Avda.
Calvo Sotelo 18, E-33007 Oviedo, Asturias, Spain}

\ead{joaquin.diaz@obspm.fr ; diego.rubiera-garcia@obspm.fr}

\begin{abstract}
We summarize the main features of a class of \emph{asymptotically anomalous} (asymptotically flat, but non Schwarzschild-like) gravitational configurations in models of gravitating non-linear electrodynamics in three space dimensions, whose Lagrangian densities are defined as arbitrary functions of the two field invariants and constrained by several physical admissibility conditions. This class of models and their associated electrostatic spherically symmetric black hole solutions are characterized by the behaviours of the Lagrangian densities around the vacuum and at the boundary of their domain of definition.
\end{abstract}

\section{Preliminaries}
For several decades a great deal of attention has been focussed on some gravitating non-linear electrodynamics (G-NED) as generalizations of the Reissner-Nordstr\"om (RN) solution of the Einstein-Maxwell field equations, whose more popular example is the gravitating Born-Infeld (BI) theory \cite{BI}. Some other G-NED models supporting asymptotically Schwarzschild-like solutions have been also considered in the literature \cite{NED}. In addition to such classes of models (and gravitational configurations) there are others supporting electrostatic spherically symmetric black hole (ESSBH) solutions approaching asymptotic flatness slower than the Schwarzschild field \cite{anomalous}. In this work we shall analyze the full class of physically admissible G-NEDs supporting these kind of \emph{asymptotically anomalous} solutions as well as their generalizations to non-abelian cases.

The action determining the dynamics of a G-NED is defined as

\begin{equation}
S=S_{G} + S_{NED} = \int d^4x \sqrt{-g}\left[\frac{R}{16\pi G} - \varphi(X,Y)\right],
\label{lagrangian}
\end{equation}
where $\varphi(X,Y)$ is a (unspecified) function of the two field invariants $X=-\frac{1}{2}F_{\mu\nu}F^{\mu\nu}=\vec{E}^2-\vec{H}^2, Y=-\frac{1}{2}F_{\mu\nu}F^{*\mu\nu}=2\vec{E}\cdot \vec{H}$, which are built from the field strength tensor $F_{\mu\nu}$ and its dual $F_{\mu\nu}^{*}=\frac{1}{2}\epsilon_{\mu\nu\alpha\beta}F^{\alpha\beta}$ and where the electric $\vec{E}$ and magnetic $\vec{H}$ fields are defined in the usual way. The associated energy-momentum tensor in flat space is obtained as

\begin{equation}
T_{\mu\nu} = 2F_{\mu\alpha}\left(\frac{\partial \varphi}{\partial X}F_{\nu}^{\alpha} + \frac{\partial \varphi}{\partial Y} F_{\nu}^{*\alpha}\right)- \varphi(X,Y) \eta_{\mu\nu}.
\end{equation}
Consistently with previous conventions we shall call \emph{physically admissible} the models satisfying the minimal requirements of (i) definiteness, continuity and regularity of $\varphi(X,Y)$ in its domain of definition of the $X-Y$ plane, assumed to be open, connected and including the vacuum, (ii) parity invariance $\varphi(X,Y)=\varphi(X,-Y)$, (iii) positive definite character of the energy, endorsed by the necessary and sufficient condition

\begin{equation}
\rho \geq \left(\sqrt{X^{2}+Y^{2}} + X\right) \frac{\partial \varphi}{\partial X} + Y \frac{\partial \varphi}{\partial Y} - \varphi(X,Y) \geq 0.
\label{energy}
\end{equation}
The associated \textit{flat-space} field equations lead, for the electrostatic spherically symmetric (ESS) solutions ($E(\vec{r}) = E(r)\vec{r}/r, \vec{H}=0$), to the first-integral

\begin{equation}
r^{2} \frac{\partial \varphi}{\partial X}\vert_{Y=0} E(r) = Q,
\label{FI}
\end{equation}
where $Q$ is an integration constant, identified as the electric charge. This equation allows the complete determination of the ESS solutions for admissible models. Assuming a field behaviour of the form $E(r \rightarrow 0)\sim r^p$ we obtain three possible cases: ultraviolet divergent (\textbf{UVD}) when $p\leq -1$, for which the integral of the energy diverges around $r=0$; \textbf{A1} when $-1<p<0$, for which the field diverges at $r=0$ but the integral of energy converges there; \textbf{A2} when $p=0$ (and $E(r \rightarrow 0) \sim a-br^{\sigma}$ with $\sigma>0$) for which the central field is finite and the integral of energy converges there. Concerning the asymptotic behaviour of the solutions, assumed to be of the form $E(r\rightarrow \infty) \sim r^q$, we are lead to four possible cases: infrared divergent (\textbf{IRD}) when $-1\leq q<0$, for which the fields vanish as $r \rightarrow \infty$ but the integral of energy diverges there; \textbf{B1} ($-2 < q < -1$), \textbf{B2} ($q = -2$) and \textbf{B3} ($q < -2$) cases, for which the integral of energy converges asymptotically. The twelve combinations of these central and asymptotic behaviours exhaust the full class of admissible NEDs. Excluding the models leading to UVD and IRD behaviours we obtain six classes exhausting the set of admissible models supporting finite-energy ESS solutions. The minimal coupling to gravity of the nine classes of models obtained by excluding the IRD behaviour was extensively analyzed in Ref.\cite{dr10a}. This analysis determines and classifies the possible G-ESS configurations, which are reduced to \textit{extreme BHs} (EBH), \textit{one or two-horizons BHs}, \textit{extreme and non-extreme black points} and \textit{naked singularities} (NS), depending on the values of $Q$ and the ADM mass coming from the integration of the field equations. All these configurations behave asymptotically as the Schwarzschild field. As we shall see, the three classes of models supporting ESS solutions with IRD behaviour lead to asymptotically anomalous gravitational configurations whose structures at finite $r$ are similar to those with normal asymptotic behaviour.

\section{Asymptotically anomalous gravitational configurations}

Let us write the general expression of the G-ESS solutions associated to the action (\ref{lagrangian}). The symmetry of the source $T_0^0=T_1^1$ leads to the Schwarzschild-like form of the static spherically symmetric line element \cite{dr10a}

\begin{equation}
ds^2=\lambda(r)dt^2-\lambda^{-1}(r)dr^2-r^2 d \Omega^2, \label{metric0}
\end{equation}
where $d\Omega^2=d\theta^2+ \sin^2 \theta d\varphi^2$. With the line element (\ref{metric0}) both the first-integral (\ref{FI}) and $X=\vec{E}^2(r,Q)$ have the same expression in flat space and in the gravitational context, allowing an immediate translation of all results obtained in the former case to the latter one.

Let us analyze three classes of admissible models supporting asymptotically IRD solutions with central field behaviours A1, A2 or UVD. Now the integral of energy diverges at $r \rightarrow \infty$ as $r^{q+1}, (-1 \leq q < 0)$ but converges at the center in the A1 and A2 cases, allowing the definition of the \emph{interior integral of energy} $\varepsilon_{in}(r,Q) = \int_0^r R^2T_0^0(R,Q)dR$ which, owing to the admissibility conditions, is a \textbf{monotonically increasing and convex} function of $r$. The general solution of the Einstein equations for the metric reads in these cases

\begin{equation}
\lambda(r,Q,C)=1+\frac{C}{r}-\frac{2 \varepsilon_{in}(r,Q)}{r},
\label{metric1}
\end{equation}
where $C$ is an integration constant. The radii of the horizons (zeroes of (\ref{metric1})) are obtained as solutions of $\frac{r+C}{2}=\varepsilon_{in}(r,Q)$ and thus are given by the intersections between the curves $\varepsilon_{in}(r,Q)$ and the beam of straight lines $(r+C)/2$ corresponding to different values of $C$ and $Q$ (see Fig.1). Through this procedure we are led to the following structures for the configurations:

\begin{figure}[]
\begin{center}
\includegraphics[width=9cm,height=4.15cm]{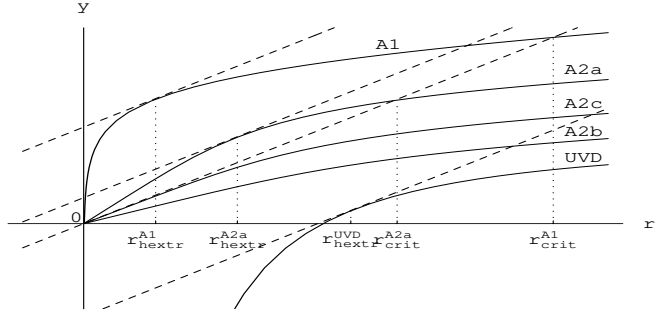}
\caption{\label{fig:1}: Interior integral of energy for the different families of NED models with asymptotically IRD field behaviours, cut by the beam of straight lines $(r+C)/2$.}
\end{center}
\end{figure}

\subsection{A1-IRD}

In this case $\varepsilon_{in}(r,Q)$ has a divergent slope as $r \sim 0$ and the above method leads to five classes of configurations. In particular there is a value $C=C_{extr}(Q)$, corresponding to the straight line tangent to the curve $\varepsilon_{in}(r,Q)$, defining the radius $r_{hextr}(Q)$ of a EBH configuration (for which the horizon is degenerate, i.e. $\lambda(r_{h})=\lambda'(r)\vert_{r=r_h}=0$). These values are obtained as solutions of

\begin{equation}
8\pi r_{hextr}^2(Q)T_0^0(r_{hextr},Q)=1\hspace{0.2cm};\hspace{0.2cm} C_{extr}(Q)= 2 \varepsilon_{in}(r_{hextr}(Q),Q) - r_{hextr}(Q).
\label{cextr}
\end{equation}
As easily seen in Fig.1, for $C>C_{extr}(Q)$ there are not cut points and the associated configurations are NS, while for $0<C<C_{extr}(Q)$ they are two-horizons BHs. On the other hand, for $C<0$ there is always a single cut point corresponding to BHs with a single (non-extreme) horizon. In the critical case ($C=0$) the metric diverges to $-\infty$ around $r = 0$ as

\begin{equation}
\lambda(r)=1 + \frac{32\pi Q}{(2-p)(p+1)}r^p +\cdot \cdot \cdot,
\label{asymp}
\end{equation}
and increases monotonically to $\lambda(r \rightarrow \infty) = 1$ exhibiting a single-horizon as well. Consequently the central singularity is timelike for $C>0$ and spacelike for $C\leq 0$. As $C \rightarrow 0^{\pm}$ in Eq.(\ref{metric1}) the metric function converges to the critical one for $r > 0$ but differ at the center by $\delta$-like terms. At large $r$, for all these IRD cases the asymptotic behaviour of the metric (\ref{asymp}) approaches flatness as $\lambda(r \rightarrow \infty) - 1 \sim -r^{q}$ (for $-1 < q < 0$, and as $\sim \ln(r)/r$ for $q=-1$) and is not Schwarzschild-like. Thus the ADM mass is not defined in the asymptotically ``anomalous" cases.

\subsection{A2-IRD}

From Fig.1 we see that similar configurations as in case A1 are obtained for $C \neq 0$ while for the critical configuration, $C=0$, the metric at the center (for $\sigma\neq 2$), which behaves as

\begin{equation}
\lambda(r \rightarrow 0,C=0,Q,a,b,\sigma) \sim 1 - 16\pi Qa + \frac{32\pi bQ}{(\sigma+1)(2-\sigma)}r^{\sigma}+ \chi r^{2},
\end{equation}
($\chi$ an integration constant) is finite. Its sign there depends on the quantity $16 \pi Q a = Q/Q_{c}$. In the \textbf{A2a} case the critical metric takes the value $\lambda(0)=1-Q/Q_{c}>0$ and its slope is

\begin{equation}
\frac{\partial \lambda}{\partial r} \Big \vert_{r \rightarrow 0} \sim \frac{32\pi b Q \sigma}{(\sigma+1)(2-\sigma)} r^{\sigma-1}+ 2\chi r.
\end{equation}
This implies that the slope of the metric at the center for the critical configuration is non-negative and depends on $\sigma$, becoming divergent for $\sigma<1$, vanishing for $\sigma>1$ and taking a finite value for $\sigma=1$. As a consequence the critical metric, which is always monotonically increasing, leads to time-like NS. On the other hand, in case \textbf{A2b} corresponding to $\lambda(0)=1-16 \pi Qa<0$ the critical configuration exhibits a single-horizon BH with a spacelike singularity at the center. The case \textbf{A2c} with $\lambda(0)=1-16 \pi Qa=0$ leads to extreme black points \cite{dr10b}.

\subsection{UVD-IRD}

In this case $\varepsilon_{in}(r,Q)$ cannot be defined, but now the Einstein equations can be integrated as
\begin{equation}
\lambda(r,Q,D) = 1 + \frac{D}{r} - \frac{2\varepsilon(r,Q,0)}{r},
\end{equation}
where $\varepsilon(r,Q,\Gamma) = 4\pi \int r^{2} T_{0}^{0}(r,Q) + \Gamma$ and the arbitrary constant $\Gamma$ has been absorbed in the constant $D$. This function is monotonically increasing and convex, exhibiting a vertical asymptote at $r=0$ and diverging with vanishing slope as $r \rightarrow \infty$. Following the same procedure as in the previous cases, the radii of the horizons are obtained from the cutting points of the function $\varepsilon(r,Q,0)$ with the beam of straight lines $(r+D)/2$, corresponding to different values of $D$. There is again a tangent line to the curve $\varepsilon(r,Q,0)$, determining the radius of an EBH configuration through the same equation (\ref{cextr}). For $D>D_{hextr}(Q)$ the configurations are NS, while two-horizons BHs appears for $D<D_{hextr}(Q)$, as can be seen from Fig.1.

\section{Discussion}

The analysis of the asymptotically anomalous class of NED models considered here, together with the one of the asymptotically Schwarzschild-like class studied in \cite{dr10a}, exhausts the family of G-ESS solutions of physically admissible NEDs. For these families, aside from NS, EBHs and two-horizon BHs, already present in the RN case, there are in addition single-horizon BHs and (in case A2 models) extreme and non-extreme black points \cite{dr10b}. Thermodynamics of the gravitating ESS structures (for the asymptotically Schwarzschild-like cases) can be consistently stated, making use of the fact that the elementary solutions of such NED models satisfy both the zeroth and first laws of BH thermodynamics \cite{thermo}, is currently in progress.

Let us finally point out that the present analysis can be extended to non-abelian gauge fields. In these cases the lagrangian densities are still given by (\ref{lagrangian}), where the field invariants are obtained by using the usual prescription in the calculation of the traces over the gauge group generators, leading to $X = -\frac{1}{2} \Sigma_{a}F_{\mu\nu a}F^{\mu\nu}_a = \Sigma_{a}(\vec{E}_a^2-\vec{H}_a^2); Y = -\frac{1}{2} \Sigma_{a}F_{\mu\nu a}F^{*\mu\nu}_a = 2\Sigma_{a}\vec{E}_a\cdot \vec{H}_a$, $a=1\cdot \cdot \cdot N$ (however, other prescriptions are possible). This problem can be reduced, for ESS fields ($\vec{E}_{a}(r)\neq 0, \vec{H}_{a}=0$), to the abelian one in a correspondence between ESS fields of both theories associated to the same form of the Lagrangian density, solving the gravitating problem for the former in terms of the solution of the latter. This extension will appear in a forthcoming publication.

\end{document}